\documentclass[12pt,a4paper]{article}
\usepackage[english]{babel}
\pdfoutput=1
\usepackage[utf8]{inputenc}
\usepackage{amsmath,amssymb}
\usepackage{float}
\usepackage{hyperref}

\newcommand\be{\begin{equation}}
\newcommand\ee{\end{equation}}
\newcommand\bea{\begin{eqnarray}}
\newcommand\eea{\end{eqnarray}}

\begin{document}

\begin{titlepage}

\vspace*{1.0cm}

\begin{center}
{\textbf{\large  4D Spherically Symmetric Time-Dependent Quantum Gravity Amplitudes 
}}
\end{center}
\vspace{1.0cm}

\centerline{
\textsc{\large J. A.  Rosabal}
\footnote{jarosabal80@gmail.com}
}

\vspace{0.6cm}

\begin{center}
{\it Departamento de Electromagnetismo y Electr{\'o}nica, Universidad de Murcia,
Campus de Espinardo, 30100 Murcia, Spain.}\\
\end{center}

\vspace*{1cm}

\begin{abstract}

In these short notes, we compute non-perturbatively the time-dependent quantum gravity amplitudes for a four-dimensional spherically symmetric space-time with space-like and time-like boundaries. We solve the 4D classical and quantum constraints in a novel way. We identify the classical solution of the constraints as a canonical transformation, where the integration constants are the new variables. We apply this canonical transformation to the path integral representation of the amplitudes we are interested in. This procedure allows us to get the time-dependent amplitudes from the path integral without solving it explicitly. From these amplitudes, we show that for most of the boundary conditions time evolution in quantum gravity is non-unitary. There is however a special case where unitary evolution could be achieved. 

\end{abstract}

\thispagestyle{empty}
\end{titlepage}

\setcounter{footnote}{0}

\tableofcontents

\newpage

\section{Introduction}\label{sec:1}

Since the seminal works \cite{Thiemann:1992jj,Kastrup:1993br,Thiemann:1993qh,Kuchar:1994zk} spherically symmetric quantum gravity (QG)  has become a paradigm. They showed that this system is fully quantizable in a  non-perturbative way. References \cite{Thiemann:1992jj,Kastrup:1993br,Thiemann:1993qh} deal with the problem of gravity quantization using Asthekar variables \cite{Ashtekar:1986yd}, which provide a straightforward way for solving the constraint. In reference \cite{Kuchar:1994zk} The main ingredient is a canonical transformation which reduces the action to the true degree of freedom. In this last reference, the GQ constraints are also solved, however, it is not an easy task. 

On the other hand in two dimensions there is a straightforward procedure to solve the constraints \cite{Henneaux:1985nw}. This method, as well as the previously described, applies to open or compact spaces but not to spaces with spatial boundaries. The inclusion of spatial boundaries opens the windows to incorporate time in a QG description.  

Recently \cite{Rosabal:2021fao,Rosabal:2023kdg} there has been some advance in the study of QG on manifold with spatial boundaries together with boundaries in time. These kinds of manifolds are of special interest because they provide a natural framework to compute time-dependent QG amplitudes. For early works on this subject see \cite{Hayward:1992ix,Hayward:1993my,Lau:1995fr,Kummer:1996si, Hawking:1996ww,Hawking:1995fd}.

This work aims to extend the results of \cite{Rosabal:2023kdg} to four dimensions. In particular, we are interested in the quantum gravity time-dependent amplitudes for the spherically symmetric space-time. Our results are not limited to gravity on these particular bounded manifolds. Our results could also be applied to open and compact spaces with this symmetry. However, for these cases, no physical time appears in the description, so no time evolution is available. 

The main motivation to compute these amplitudes is because through them we can diagnose whether or not the time evolution of a QG system is unitary, without having to compute the time-dependent entropy of the system \cite{Hawking:1976ra}.

In  \cite{Rosabal:2023kdg}, a recipe to compute the time-dependent quantum gravity amplitudes in two dimensions was used. Although in this reference it was not explicitly given, here we list its steps.
\begin{enumerate}
\item Find a classical solution to the Hamiltonian and the Momentum constraints in the form (if possible) $\text{P}_{\text{A}}=\text{M}\big[\text{Q}^{\text{A}}\big]$. Where $\big(\text{Q}^{\text{A}},\text{P}_{\text{A}}\big)$, represent the coordinates and the momenta of the theory.
\item Find a quantum solution to the constraints ($\text{P}_{\text{A}}\rightarrow-\text{i}\frac{\delta{}}{\delta \text{Q}^{\text{A}}}$) in the form $\Psi=\text{exp}[\text{i}\Omega]$.
\item Promote the classical solution to a change of variable in the path integral,  the new variables will be the integration constants associated with this solution. By construction, it is a canonical transformation.
\item Identify $\Omega$ as the generating function of the transformation.
\item Perform this transformation to the path integral.
\item Solve for the new theory.
\item Get the amplitudes.
\end{enumerate}

Although this procedure applies to open or compact spaces, here we are interested in space-times with spatial boundaries. To account for them we need to complement the previous list with one more step before getting the amplitudes. This is
\begin{itemize}
\item Reduce the boundary action to be able to apply the canonical transformation. 
\end{itemize}
This step can be done always, for any gravity theory on a manifold with spatial boundaries. See for instance the form of the boundary action in \cite{Rosabal:2021fao} for more general settings beyond spherically symmetric space-times. What we do not know is whether the reduced action is always adapted to the canonical transformation, if it exists. For sure for the two \cite{Rosabal:2023kdg}, and four-dimensional (as we will see shortly) cases the reduced boundary action is perfectly adapted to be transformed by the canonical transformation associated with the classical solution to the constraints.

In these short notes, we will apply this recipe to the calculation of the 4D time-dependent QG amplitudes in the spherically symmetric space-time with spatial boundaries. The key ingredient will be the use of the same method proposed by Marc Henneaux in \cite{Henneaux:1985nw}, and refined by D. Louis-Martinez et al. in \cite{ Louis-Martinez:1993bge} to solve the Hamiltonian and the Momentum constraints but in our case for the four-dimensional Hamiltonian, which differs from the two dimensional one. Once this solution is found the rest of the steps can be easily performed, almost mimicking the two-dimensional case \cite{Rosabal:2023kdg}. The second fundamental ingredient is a canonical transformation that allows to reduce the gravity action in the same spirit of \cite{Kuchar:1994zk}. The novelty of this transformation is that it can be constructed directly from the classical solution of the constraints.

The paper is organized as follows, in section  \ref{sec:2} we briefly review the canonical formalism for spherically symmetric space-time with space-like and time-like boundaries.  We present a new procedure to solve the constraint for this case. This approach is inspired by the method presented in \cite{Henneaux:1985nw} for the two-dimensional case. 

In section \ref{sec:3} we develop the QG theory of the spherically symmetric space-time with spatial boundaries together with boundaries in time. We solve the quantum version of the constraints. After,  we identify the classical and quantum solutions of the constraints with the canonical transformation and the generating function of the transformation, respectively. Finally, we present the time-dependent amplitude. Interestingly enough for most of the boundary conditions, the amplitude develops a non-unitary behavior.  Nonetheless, there is a very special case in which the amplitude could evolve unitary. Conclusions are presented in section \ref{conslusion} followed by three appendices. Two of these appendices are devoted to working out the changes of variables in the path integral. The third concerns the conditions for the wave function to evolve unitary.

\section{Canonical formalism for spherically symmetric space-time with space-like and time-like boundaries}\label{sec:2}

The canonical formalism for spherically symmetric (open) space-time was developed in \cite{Kuchar:1994zk}. The introduction of spatial boundaries together with the proper boundary terms was presented in \cite{Lau:1995fr}.
The formalism for more general space-times, beyond those with spherical symmetry, has been developed in \cite{Hayward:1993my, Hayward:1992ix,Hawking:1996ww, Rosabal:2021fao,Hawking:1995fd}. Here we briefly review it using a more modern notation. We also present a new procedure to find the classical solution of the constraints in four dimensions. Although in two dimensions a similar procedure has been used and extensively celebrated \cite{Henneaux:1985nw} in four dimensions, up to the knowledge of the author, it has never been explored.

The ADM metric for the spherically symmetric space-time  reads like 
\be
  \text{ds}^2  =-\text{N}^2\text{dt}^2 +\Lambda^2 (\text{dr}+\text{N}^{\text{r}}\text{dt})^2+4\phi \ \text{ds}^2_{\text{S}^2}.
\ee
We have used $\phi$ to contrast more easily with the two-dimensional case \cite{Rosabal:2023kdg}.
The Einstein Hilbert action in the Hamiltonian form on a spherically symmetric space-time  with spatial boundaries takes the form 
\begin{align}\label{action}
\text{S}=\text{S}_{\text{M}}+\text{S}_{\text{B}}= &\int\limits_{t_i}^{t_f}\text{d}t\int\limits_{0}^{\text{r}_0}\text{dr}\Big[ \text{P}_{\phi}\partial_t\phi+\text{P}_{\Lambda}\partial_t \Lambda-\text{N}\cal{H}
-\text{N}^{\text{r}}\cal{H}_{\text{r}}\Big]\\ \nonumber
{} & +\int\limits_{\text{B}}\text{d}t \Big(\Lambda\text{N}^{\text{r}}\ \text{P}_{\Lambda}-2\text{N}\Lambda^{-1}\partial_{\text{r}}\phi+2\partial_t\phi \ \text{arcsinh}\big(\frac{\Lambda\text{N}^{\text{r}}}{\overline{\text{N}}} \big) \Big).
\end{align}
Where
\be 
{\cal H}  = -\frac{1}{2}\text{P}_{\phi}\text{P}_{\Lambda}+2\partial_{\text{r}}\big(\Lambda^{-1}\partial_{\text{r}}\phi\big)-\frac{1}{2}\Lambda\ +\frac{1}{8}\Lambda\phi^{-1}\Big(\text{P}_{\Lambda}^2-(2\Lambda^{-1}\partial_{\text{r}}\phi)^2 \Big),\label{HC}
\ee
\be
{\cal H}_{\text{r}}  =  \partial_{\text{r}}\phi\text{P}_{\phi}-\Lambda\partial_{\text{r}}\text{P}_{\Lambda}\label{MC}.
\ee

The fields in the action \eqref{action} are subjected to the following boundary conditions. On the initial and final slices we fix $(\Lambda, \phi)_i$, and $(\Lambda, \phi)_f$, respectively. The induced metric on the spatial boundary, at a constant $\text{r}=\text{r}_0$, takes the form
\be
  \text{ds}^2_{\big{|}_{\text{r}_0}}  =-(\text{N}^2-(\Lambda\text{N}^{\text{r}})^2)\text{dt}^2+4\phi \ \text{ds}^2_{\text{S}^2}= -\overline{\text{N}}^2\text{dt}^2+4\phi \ \text{ds}^2_{\text{S}^2}.
\ee
On this boundary the combination 
\be\label{BC}
\text{N}^2-(\Lambda\text{N}^{\text{r}})^2= \overline{\text{N}}^2,
\ee
is fixed, with $\overline{\text{N}}$, a given function depending only on time. So, over $\text{B}$, we fix 
\be
\big(\sqrt{\text{N}^2-(\Lambda\text{N}^{\text{r}})^2},\phi\big)_{\big{|}_{\text{r}_0}}.
\ee

Classically, variations with respect to $\text{N}$, and  $\text{N}^{\text{r}}$,  in the bulk, impose the constraint 
\be
{\cal H}={\cal H}_{\text{r}}=0.
\ee
 Note that over the spatial boundary $\text{N}$, and  $\text{N}^{\text{r}}$, are not independent \eqref{BC}, so extra care is needed with these variations over the time-like boundary.
   
In the same spirit \footnote{Note that in \cite{Henneaux:1985nw}, and in general for all two dimensional dilatonic gravities \cite{Louis-Martinez:1993bge}, the Hamiltonian corresponds with the three first terms in \eqref{HC}.} of \cite{Henneaux:1985nw} we can make the linear combination
\be
{\cal H}\partial_{\text{r}}\phi+\frac{1}{2}{\cal H}_{\text{r}}\text{P}_{\Lambda}=0\nonumber,
\ee
to get
\be
-\partial_{\text{r}}\Big(\text{P}_{\Lambda}^2-(2\Lambda^{-1}\partial_{\text{r}}\phi)^2 \Big)+\frac{1}{2}\phi^{-1}\partial_{\text{r}}\phi\Big(\text{P}_{\Lambda}^2-(2\Lambda^{-1}\partial_{\text{r}}\phi)^2 \Big)-2\partial_{\text{r}}\phi=0.
\ee
Previous relation is a differential equation for the function  
\be
\text{F}=\text{P}_{\Lambda}^2-(2\Lambda^{-1}\partial_{\text{r}}\phi)^2,
\ee
\be\label{eqF}
-\partial_{\text{r}}\text{F}+\frac{1}{2}\phi^{-1}\partial_{\text{r}}\phi \text{F}-2\partial_{\text{r}}\phi=0.
\ee

The particular form of this equation allows us to solve it in terms of $\phi$, solely
\be\label{equdifgood}
\text{F}=\text{C}(t)\sqrt{\phi}-4\phi=\text{P}_{\Lambda}^2-(2\Lambda^{-1}\partial_{\text{r}}\phi)^2.
\ee
To finally get the full solution of the constraints as
\bea\label{classsolutioncontr}
\text{P}_{\Lambda} & = & \Big(\text{F}+(2\Lambda
   ^{-1}\partial_{\text{r}}\phi)^2 \Big)^{\frac{1}{2}},\nonumber\\
\text{P}_{\phi} & = &\frac{\Lambda}{\partial_r \phi}\partial_{\text{r}} \text{P}_{\Lambda}=\frac{\Lambda}{\partial_r \phi}\frac{\partial_r\Big( \text{F}+
    (2\Lambda
   ^{-1}\partial_{\text{r}}\phi)^2\Big)}{2\Big(\text{F}+(2\Lambda
   ^{-1}\partial_{\text{r}}\phi)^2\Big)^{\frac{1}{2}}}.
\eea

  Note that  \eqref{classsolutioncontr} written in terms of $\text{F}$ can be generalized through the inclusion of more inhomogeneous terms depending on $\phi$ and its derivatives. 
\be\label{eqF-gene}
-\partial_{\text{r}}\text{F}+\frac{1}{2}\phi^{-1}\partial_{\text{r}}\phi \text{F}-2\partial_{\text{r}}\phi=\text{W}[\phi].
\ee  
This generalization in four dimensions is equivalent to the statement that all dilatonic gravities in two dimensions are solvable \cite{Louis-Martinez:1993bge}.  In four dimensions, however,  extra care is needed because $\phi$, is not a scalar but a component of the metric. This means that $\text{W}[\phi]$, should come from an invariant term in the action. For instance, a cosmological constant term in four dimensions is allowed. In this case, in the action, we would have an extra term
\be
\lambda \sqrt{-g}=\text{N}\Lambda (4\lambda\phi)\Rightarrow \text{W}[\phi]\propto \lambda\phi\partial_{\text{r}}\phi=\frac{1}{2} \lambda \ \partial_{\text{r}}\phi^2.
\ee
Extensions of  \text{W}, to include  $\Lambda$, and their derivatives are possible but they are out of the scope of this work.

\section{Spherically symmetric QG with space-like and time-like boundaries}\label{sec:3}

In the previous section, we have performed the step one of the recipe to get the amplitudes we are interested in. In this section, we perform the rest of the steps on the list. 

To complete the second point we need to solve the functional differential equations 
\bea\label{Quantumsolutioncontr}
-\text{i}\frac{\delta}{\delta\Lambda}\Psi & = & \Big(\text{F}+(2\Lambda
   ^{-1}\partial_{\text{r}}\phi)^2 \Big)^{\frac{1}{2}}\Psi, \nonumber\\
-\text{i}\frac{\delta}{\delta\phi}\Psi & = &\frac{\Lambda}{\partial_r \phi}\frac{\partial_r\Big( \text{F}+
    (2\Lambda
   ^{-1}\partial_{\text{r}}\phi)^2\Big)}{2\Big(\text{F}+(2\Lambda
   ^{-1}\partial_{\text{r}}\phi)^2\Big)^{\frac{1}{2}}}\Psi.
\eea

These equations are equivalent to the Wheeler-DeWitt equation and the Momentum constraint. Their integration is similar to the two-dimensional case \cite{Henneaux:1985nw, Louis-Martinez:1993bge, Rosabal:2023kdg}
\be\label{notamplitude}
\Psi= \text{exp}\Big[\text{i} \ \Omega^{\prime}[\Lambda,\phi;\text{C}] \Big],
\ee
with
\be
\Omega^{\prime}[\Lambda,\phi;\text{C}]=\Omega[\Lambda,\phi;\text{C}]+\text{G}(\text{C})\label{gene-func0},
\ee
\begin{align}\nonumber
\Omega[\Lambda,\phi;\text{C}]= & \int\limits_{\Sigma_{t}}\text{dr}\  \Lambda\Big[\sqrt{\text{F}+(2\Lambda
   ^{-1}\partial_{\text{r}}\phi)^2}\\
   {} & 
   +(2\Lambda^{-1}\partial_{\text{r}}\phi)
   \text{ln}\Big( \frac{(2\Lambda^{-1}\partial_{\text{r}}\phi)+\sqrt{\text{F}+(2\Lambda
   ^{-1}\partial_{\text{r}}\phi)^2}}{(2\Lambda^{-1}\partial_{\text{r}}\phi)-\sqrt{\text{F}+(2\Lambda
   ^{-1}\partial_{\text{r}}\phi)^2}} \Big)
    \Big]\label{gene-func1},
\end{align}
where $\text{G}(\text{C})$, is a real arbitrary function of the constant of integration $\text{C}(t)$.

If we were computing the wave function on a compact or an open manifold \eqref{notamplitude} would be the final solution. Note however that there is no room for introducing time.

Let us now pose the problem on the corresponding space-time with boundaries. We are interested in computing the transition amplitude between an initial configuration $(\Lambda, \phi)_{i}$, at some initial time $t_i$,  and a final one $(\Lambda,\phi)_{f}$, at some final time $t_f$. While over $\text{B}$ the configuration is $(\overline{\text{N}},\phi)_{\text{B}}$.

In the path integral formulation, this transition amplitude is represented by
\begin{align}\label{T1}
\Psi\Big[(\Lambda,\phi)_{f},(\Lambda,\phi)_{i}\ ;(\overline{\text{N}},\phi)_{\text{B}}\Big]  & =\int\text{D}\big[\text{N},\text{N}^{\text{r}},\phi,\Lambda,\text{P}_{\phi}, \text{P}_{\Lambda}\big]_{\big{|}_{\text{M}}}\nonumber \\
{} & \ \ \ \int\text{D}\big[\text{N},\text{N}^{\text{r}},\Lambda, \text{P}_{\Lambda}\big]_{\big{|}_{\text{B}}}\nonumber\\
{} & \delta\big[ \text{N}^2 -(\Lambda \text{N}^{\text{r}} )^2-\overline{\text{N}}^2\big]_{\big{|}_{\text{B}}}
\text{e}^{\text{i}\text{S}}.
\end{align}
Where the measure $\text{D}\big[\ldots \big]$, might depend on $(\phi,\Lambda)$, but it does not depend on the momenta $(\text{P}_{\phi},\text{P}_{\Lambda})$, nor on the Lagrange multipliers $(\text{N},\text{N}^{\text{r}})$, except for the functional Dirac delta. Note that after integration in $(\text{P}_{\phi},\text{P}_{\Lambda})$, we can recover the Einstein-Hilbert action in its Lagrangian form. See \cite{Kuchar:1994zk}, for the Lagrangian form of the action. 

Performing the change of variables over the spatial boundaries only
\bea\label{change}
\Lambda \text{N}^{\text{r}} & = & {\cal R}\ \text{sinh}(\eta),\\
\text{N} & = & {\cal R} \ \text{cosh}(\eta) \nonumber,
\eea
where ${\cal R}$, ranges from $[0,\infty)$, while  $\eta$, ranges in the interval $(-\infty,\infty)$.
The transition amplitude can be written as,  appendix \ref{appA}
\be\label{T2}
\Psi\Big[(\Lambda,\phi)_{f},(\Lambda,\phi)_{i}\ ;(\overline{\text{N}},\phi)_{\text{B}}\Big]   =\int\text{D}\big[\text{N},\text{N}^{\text{r}},\phi,\Lambda,\text{P}_{\phi}, \text{P}_{\Lambda}\big]_{\big{|}_{\text{M}}} \int\text{D}\big[\eta,\Lambda, \text{P}_{\Lambda}\big]_{\big{|}_{\text{B}}}
\text{e}^{\text{i}\text{S}}.
\ee
Where now
\be
S_{\text{B}}=\int\limits_{\text{B}}\text{d}t \Big(\overline{\text{N}}\ \text{P}_{\Lambda}\ \text{sinh}(\eta)-2\overline{\text{N}}\Lambda^{-1}\partial_{\text{r}} \phi\  \text{cosh}(\eta)+2\partial_t \phi\eta \Big).
\ee

Proceeding with the last complementary point on the list we can reduce the boundary action. To do that we should note first that the measure after the change of variables does not depend on $\eta$, see appendix \ref{appA}. Second, the equation of motion for $\eta$, does not contain derivatives. This fact allows us to solve for $\eta$, in terms of the other boundary degree of freedom and plug back the solution into the action, to get \cite{Rosabal:2023kdg}
\be\label{B...action...}
\text{S}_{\text{B}}  =  -\int\limits_{\text{B}}\text{d}t\overline{\text{N}}\sqrt{(2\Lambda
   ^{-1}\partial_{\text{r}}\phi)^2- \text{P}_{\Lambda}^2}.
\ee
We have assumed that 
\be
\phi_{|_{\text{B}}}=\text{const}\Rightarrow \partial_t\phi_{|_{\text{B}}}=0.
\ee
See also \cite{Rosabal:2023kdg,Lau:1995fr,Kummer:1996si}, where a similar treatment of the boundary action is performed.

After checking points 3 and 4 on the list we are in a condition of performing the change of variables \eqref{classsolutioncontr} in the path integral \eqref{T2}. Note that \eqref{classsolutioncontr} is the transformation that maps ${\cal H}$,  and ${\cal H}_{\text{r}}$, to zero. So, we expect the action
gets reduced considerably. In addition, it is a canonical transformation, therefore the new action will contain two boundary terms at the initial and final time. 

After performing the canonical transformation we get
\begin{align}\label{actfinalformcanno}
\text{S} & =\int\limits_{t_i}^{t_f}\text{d}t\int\limits_{0}^{\text{r}_0}\text{dr}\Big[ \text{P}_{\text{C}}\partial_t\text{C}+\tilde{\text{P}}_{\phi}\partial_t \phi-{\cal K}[\text{C},\phi,\text{P}_{\text{C}},\tilde{\text{P}}_{\phi}]\Big]\nonumber\\
   {} & +\int\limits_{t_i}^{t_f}\text{d}t\frac{\text{d}}{\text{d}t}\Omega[\Lambda,\phi;\text{C}] -\int\limits_{\text{B}}\text{d}t\ \overline{\text{N}}\sqrt{-\text{F}}.
\end{align}
where
\bea\label{cantransojo}
\text{P}_{\Lambda} & = & \frac{\delta}{\delta\Lambda}\Omega[\Lambda,\phi;\text{C}],\\
\text{P}_{\phi} & = & \tilde{\text{P}}_{\phi}+\frac{\delta}{\delta\phi}\Omega[\Lambda,\phi;\text{C}],\nonumber\\
\text{P}_{\text{C}} & = & -\frac{\partial}{\partial \text{C}}\Omega[\Lambda,\phi;\text{C}]\nonumber,
\eea
and the new Hamiltonian ${\cal K}[\text{C},\phi,\text{P}_{\text{C}},\tilde{\text{P}}_{\phi}]$,

\be
{\cal K}[\text{C},\phi,\text{P}_{\text{C}},\tilde{\text{P}}_{\phi}]  = -\frac{1}{2}\text{N}\tilde{\text{P}}_{\phi}\sqrt{\text{F}+(2\Lambda
   ^{-1}\partial_{\text{r}}\phi)^2} +\text{N}^{\text{r}}\partial_{\text{r}}\phi\tilde{\text{P}}_{\phi}.
\ee

Now integration in $\text{N}$, or $\text{N}^{\text{r}}$, imposes $\tilde{\text{P}}_{\phi}=0$, and the action further reduces to  
\be\label{complexaction}
\text{S}  =\int\limits_{t_i}^{t_f}\text{d}t  \ \Pi_{\text{C}}\ \partial_t\text{C}+\Omega[\Lambda,\phi;\text{C}]\Big{|}_{t_i}^{t_f} -\int\limits_{\text{B}}\text{d}t\ \overline{\text{N}}\sqrt{-\text{F}},
\ee
where $\Pi_{\text{C}}(t)=\int\limits_{0}^{\text{r}_0}\text{dr}\ \text{P}_{\text{C}}(t)$.

At the classical level the boundary term $\Omega[\Lambda,\phi;\text{C}]\Big{|}_{t_i}^{t_f}$, in  \eqref{complexaction} does not play any role because it does not affect the equations of motions. However, at the quantum level, as we are assuming we are performing the change of variables within the path integral we can not discard this term.

The path integral in the remaining variables $\big(\Pi_{\text{C}}(t), \text{C}(t)\big)$, and the new action can be reduced even more. As the measure does not depend on $\Pi_{\text{C}}$, see appendix \ref{appB}  the path integration in $\Pi_{\text{C}}$, implies $\text{C}(t)=\text{const}=\text{C} \ \ \forall \ \  t$, and the new action reduces to
\be\label{complexaction1}
\text{S}  =\Omega[\Lambda,\phi;\text{C}]\Big{|}_{t_i}^{t_f} -\int\limits_{\text{B}}\text{d}t\ \overline{\text{N}}\sqrt{-\text{F}},
\ee
where, now $\text{C}$, is a constant. 

Putting all these results together we finally get the spherically symmetric time-dependent amplitude   
\be\label{finalamplitudeojo}
\Psi\Big[(\phi,\Lambda)_{f},(\phi,\Lambda)_{i}\ ;(\phi,\tau)_{\text{B}}\Big]   =  \int\limits_{0}^{\infty} \text{dC}\  \chi(\text{C}) \text{exp}\Big[\text{i}\Omega[\Lambda,\phi;\text{C}]\Big{|}_{t_i}^{t_f}+ \sqrt{F}_{|_{\text{B}}} \tau\Big],
\ee

To make clear the appearance of $\chi(\text{C})$, in \eqref{finalamplitudeojo}, it is convenient to work the quantum mechanical problem derived from the action \eqref{complexaction} using the  Schr\"{o}dinger equation. For this case the Hamiltonian $\text{h}[\text{C},t]$, is given by
\be
\text{h}[\text{C},t] =\overline{\text{N}}(t)\sqrt{-\text{F}}.
\ee
The Schr\"{o}dinger equation reads
\be
\text{i}\partial_{t_f}\psi(\text{C},t_f)=\overline{\text{N}}(t)\sqrt{-\text{F}}\ \psi(\text{C},t_f).
\ee
It is straightforward to see that we can separate variables $\psi(\text{C},t_f)=\chi(\text{C}) \text{f}(t_f)$, to get 
\be
\psi(\text{C},t_f)=\chi(\text{C}) \text{exp}\Big[\int\limits_{\text{B}}\text{d}t\ \overline{\text{N}} \sqrt{\text{F}}\Big].
\ee

The function $\chi(\text{C})$, is a complex function that can be determined by imposing initial conditions for the functional $\Psi\Big[(\phi,\Lambda)_{f},(\phi,\Lambda)_{i}\ ;(\phi,\tau)_{\text{B}}\Big]$, and 
\be
\tau =  \int\limits_{t_i}^{t_f}\text{d}t \ \overline{\text{N}}.
\ee
is the proper time over the boundary. This means that in order to fully determine the function $\chi(\text{C})$, we need to prescribe the state at $\tau=0\Leftrightarrow t_i=t_f$, $\Psi\Big[(\phi,\Lambda)_{f},(\phi,\Lambda)_{i}\ ;(\phi,0)_{\text{B}}\Big]$,  namely
\be
\Psi\Big[(\phi,\Lambda)_{f},(\phi,\Lambda)_{i}\ ;(\phi,0)_{\text{B}}\Big]=\int\limits_{0}^{\infty} \text{dC}\  \chi(\text{C}) \text{exp}\Big[\text{i}\Omega[\Lambda,\phi;\text{C}]\Big{|}_{t_i}^{t_f}\Big].
\ee
Solving previous equation for $\chi(\text{C})$, complete determines the time-dependent state. 

As in the 2D case, \cite{Rosabal:2023kdg} in this example several boundary conditions for the metric on the spatial boundary will lead to a non-unitary evolution. This is determined by the sign of the function $\text{F}$. Note however that we also have room for boundary conditions leading to unitary evolution, see appendix \ref{appC}.

The amplitude \eqref{finalamplitudeojo} will potentially  evolve unitarily  only if for all values of $\text{C}$, and a given value of $\phi$, over $\text{B}$,   $\text{F}_{|_{\text{B}}}< 0$. Note that when $\text{C}<4\sqrt{\phi}$, $\text{F}< 0$,  and vice versa for the other values of $\text{C}$. This means that to have an entirely negative function for all values of $\text{C}$, we need to push to infinite the zero of $\text{F}$. In other words, 
\be 
\text{F}_{\big{|}_{\text{B}}}<0 \ \ \forall \ \ \text{C}, \ \text{only when}  \ \  \phi_{\big{|}_{\text{B}}}\rightarrow \infty.
\ee

To have a well-defined limit in the action we need to complement it with a boundary term (counterterm) evaluated on a classical reference background \cite{Hawking:1995fd}. At this point, it is worth pointing out that choosing a classical solution is equivalent to fixing a value of $\text{C}$. 

The boundary action on the reference background takes the same form as in \eqref{B...action...}. After performing the canonical transformation, or using the classical equation of motion for the reference background we have
\be
\text{S}_{\text{B}_{\text{reference}}}=-\sqrt{-\text{F}_0}_{\big{|}_{\text{B}}}\tau,
\ee
where 
\be
\text{F}_0=\text{C}_0\sqrt{\phi}-4\phi,
\ee
with $\text{C}_0$, fixed.

It is straightforward to check that 
\be 
\lim_{\phi\to\infty}\Big(\text{S}_{\text{B}}-\text{S}_{\text{B}_{\text{reference}}}\Big)=\lim_{\phi\to\infty}-\text{i}\Big(\sqrt{\text{F}}-\sqrt{\text{F}}_0\Big) \tau =\Big( \frac{1}{4}\text{C}_0- \frac{1}{4}\text{C}\Big)\tau.
\ee
So, $\text{C}_0$, enters in the wave function in the phase $\text{exp}\Big(\text{i}\frac{1}{4}\text{C}_0\tau\Big)$, namely $\text{C}_0$, does not contribute to the probability
\be\label{finalamplitudeojo1}
\Psi\Big[(\phi,\Lambda)_{f},(\phi,\Lambda)_{i}\ ;(\infty,\tau)_{\text{B}}\Big]   = \text{e}^{(\text{i}\frac{1}{4}\text{C}_0\tau)} \int\limits_{0}^{\infty} \text{dC}\  \chi(\text{C}) \text{exp}\Big[\text{i}\Omega^{\prime}[\Lambda,\phi;\text{C}]\Big{|}_{t_i}^{t_f} -\text{i}\frac{1}{4}\text{C}\tau\Big].
\ee

\section{Conclusions}{\label{conslusion}}

In this work, we have computed for the first time the time-dependent transition amplitudes for a spherically symmetric space-time \eqref{finalamplitudeojo} with space-like and time-like boundaries. From it, one can show that for most of the boundary conditions, the quantum evolution is non-unitary. There is however a case where unitary evolution could be achieved. It is when the $\phi_{|_{\text{B}}}$, component of the metric goes to infinite on the spatial boundary \eqref{finalamplitudeojo1}. In addition, there is one more condition to be satisfied. It is given in equation  \eqref{uni-ojo-ojo}. This is a very strong condition on the measure of the internal product between states. 

We do not discard the possibility that such a measure exists. Nonetheless, it is worth emphasizing that only when these two conditions are met this quantum gravity system will evolve unitary. This is a quite revealing fact since we have been struggling for decades trying to show that QG is unitary. It turns out that for most physical configurations it is not, and the windows for which it could be is very small. Certainly, finding such a measure, or showing it does not exist will end this battle for unitarity. We leave this study for future work.

Besides the results commented on in the previous paragraphs, over the paper we have derived others 
results. Perhaps the most important among them is the solutions of the classical \eqref{classsolutioncontr} and quantum constraints \eqref{Quantumsolutioncontr}. We have used a method proposed in \cite{Henneaux:1985nw} for the case of two-dimensional gravity. Its extension to four-dimensional gravity, as far as the author is concerned, is a new contribution never before explored in the literature. The power of this procedure in four dimensions and its beauty get condensed in equation \eqref{eqF} (or its generalization \eqref{eqF-gene}), and its solution \eqref{equdifgood}. 

In connection with previous results is the canonical transformation described in section \ref{sec:3}. It resembles the one presented in  \cite{{Kuchar:1994zk}}. The novelty of this transformation is that it can be identified directly with the classical solution of the constraints \eqref{classsolutioncontr}. Remarkably, the generating function, \eqref{gene-func0} and \eqref{gene-func1},  of this transformation can be derived from the quantum solution of the constraints \eqref{Quantumsolutioncontr}.

The fact that we can derive a canonical transformation from the classical solution of the constraints opens new avenues to address some QG systems. As long as we can find a classical solution to the constraints, as commented in the introduction,  the system will be non-perturbatively  quantizable. 

Certainly, an urgent point should be addressed is the inclusion of matter for the spherically symmetric space-time on this bounded manifold. With its inclusion, we will be able to incorporate in the Hilbert space states describing the black hole formation and evaporation. We leave the inclusion of matter for future works.

Before ending we would like to comment again and emphasize some points about the (non) unitary issue. Since the work of Hawking \cite{Hawking:1976ra} the physics community became aware that QG might evolve non-unitary. Since then a lot of effort has been put into showing that QG is unitary. 

In this work we have conclusively shown using the time-dependent amplitudes and not the entropy that for most of the boundary configurations, pure spherically symmetric QG in four dimensions is indeed non-unitary. The only case where it could be unitary is when $\phi_{|_{\text{B}}}\rightarrow\infty$. However, in addition, one more condition has to be satisfied to ensure unitarity \eqref{uni-ojo-ojo}. This condition is related to the measure of the internal product of the Hilbert space. 

{\it The message one should take from this work is that for most of the boundary configuration, QG evolves non-unitarily, and the case where it could be unitary is so restrictive that seems not to exist \eqref{uni-ojo-ojo}.}

\section*{Acknowledgments}
We thank Jose J. Fernandez-Melgarejo for comments and support and Marc Henneaux for the correspondence.

\appendix

\section{The boundary measure}{\label{appA}}

In this appendix, we develop the path integral boundary measure. Although it is not needed to fully know it, we need to know certain information at some points to perform some steps. We are assuming that the measure might depend on the fields degree of freedom $\text{h}=(\Lambda,\phi)$,  but it does not depend on the momenta $(\text{P}_{\phi},\text{P}_{\Lambda})$, nor on the Lagrange multipliers $(\text{N},\text{N}^{\text{r}})$. 

We start with the boundary measure in \eqref{T1}
\be
\text{D}\big[\text{N},\text{N}^{\text{r}},\Lambda, \text{P}_{\Lambda}\big]_{\big{|}_{\text{B}}}\nonumber\\
\delta\big[ \text{N}^2 -(\Lambda \text{N}^{\text{r}} )^2-\overline{\text{N}}^2\big]_{\big{|}_{\text{B}}}\label{m0}.
\ee
Given what is stated in the previous paragraph, this measure can be interpreted as 
\be\label{m1}
\prod_{t\in[t_i,t_f]}\text{H}[\text{h}] \ \text{d}\text{N}(t)\wedge \text{d}\text{N}^{\text{r}}(t) \wedge \text{d}\Lambda(t)\wedge \text{d}\text{P}_{\Lambda}(t) \delta\Big( \text{N}(t)^2 -(\Lambda(t) \text{N}^{\text{r}}(t) )^2-\overline{\text{N}}(t)^2\Big)_{\big{|}_{\text{B}}}.
\ee

Now, performing the change of variables,  the first we should note is that we can differentiate both expressions (at a fixed time) in \eqref{change} to compute the wedge product in \eqref{m1}. The first step yields to 
\be
\text{d}\Lambda \text{N}^{\text{r}}\wedge \text{d}\text{N}+\Lambda \text{d} \text{N}^{\text{r}}\wedge\text{d}\text{N}={\cal R} \text{d}\eta \wedge \text{d}{\cal R}. 
\ee
While a second step brings the product to the desirable form
\be
\Lambda \text{d}\text{N}\wedge  \text{d} \text{N}^{\text{r}} \wedge \text{d}\Lambda={\cal R} \text{d}\eta \wedge \text{d}\Lambda \wedge \text{d}{\cal R}.
\ee

The Dirac delta can be straightforwardly massaged to get 
\be
\delta\Big( {\cal R}^2-\overline{\text{N}}^2\Big)=\frac{\delta\Big( {\cal R}-\overline{\text{N}}\Big)}{2\overline{\text{N}}}+\frac{\delta\Big( {\cal R}+\overline{\text{N}}\Big)}{2\overline{\text{N}}}.
\ee
As  $\overline{\text{N}}>0$, and the integration in ${\cal R}$, is on the interval $[0,\infty)$,   only the first term will contribute to the path integral. 

Putting all the results together we can see that \eqref{m0} can be expressed as 
\be
\prod_{t\in[t_i,t_f]}\text{H}[\text{h}]_{\big{|}_{\text{B}}} \ \Lambda^{-1}(t){\cal R}(t)\ \text{d}\eta(t) \wedge \text{d}\Lambda(t) \wedge \text{d}{\cal R}(t) \ \frac{\delta\Big( {\cal R}(t)-\overline{\text{N}}(t)\Big)}{2\overline{\text{N}}(t)}.
\ee
Integration in the variable ${\cal R}$, yields to
\be
\prod_{t\in[t_i,t_f]}\frac{1}{2} \text{H}[\text{h}]_{\big{|}_{\text{B}}} \ \Lambda^{-1}(t)\ \text{d}\eta(t) \wedge \text{d}\Lambda(t) \label{m3}.
\ee
Where now the $\eta$, integration can be straightforwardly performed.

What is important to note from \eqref{m3} is that it does not depend on $\eta$. This justifies the manipulations to get \eqref{B...action...}.

\section{The bulk measure}{\label{appB}}

In this appendix, we perform the canonical transformation \eqref{classsolutioncontr} to the bulk measure 
\be
\text{D}\big[\text{N},\text{N}^{\text{r}},\phi,\Lambda,\text{P}_{\phi}, \text{P}_{\Lambda}\big]_{\big{|}_{\text{M}}}.
\ee
The interpretation of this measure is similar to the previous one but of course, the transformation is much more involved.
\be\label{mm1}
\prod_{t\in[t_i,t_f]}\prod_{\text{r}\in[0,\text{r}_0)}\text{H}[\text{h}] \ \text{d}\text{N}\wedge \text{d}\text{N}^{\text{r}} \wedge \text{d}\phi\wedge \text{d}\Lambda\wedge \text{d}\text{P}_{\phi} \wedge \text{d}\text{P}_{\Lambda}.
\ee
Under the assumptions about the measure stated in the previous section and to solve the last integration in $\big(\Pi_{\text{C}}(t), \text{C}(t)\big)$, in section $\ref{sec:3}$, it is important to show that the measure does not depend on $\Pi_{\text{C}}(t)$.

First, we can differentiate (at a fixed (t,\text{r})) the expressions in \eqref{classsolutioncontr}. The relevant parts are
\bea 
\text{d}\text{P}_{\Lambda} & = & \frac{\sqrt{\phi}}{2\big(\text{F}+(2\Lambda^{-1}\partial_{\text{r}}\phi)^2\big)^{\frac{1}{2}}}\text{d}\text{C}+\ldots \ ,\\ 
\text{d}\text{P}_{\phi} & = & \text{d}\tilde{\text{P}}_{\phi}+\ldots \ ,
\eea 
where the ellipsis represents the terms with the differentials that will cancel in the wedge product \eqref{mm1}. Plugging these expression in \eqref{mm1} we arrive at 
\be\label{mm2}
\prod_{t\in[t_i,t_f]}\prod_{\text{r}\in[0,\text{r}_0)}\text{H}[\text{h}] \ \text{d}\text{N}\wedge \text{d}\text{N}^{\text{r}} \wedge \text{d}\phi\wedge \text{d}\Lambda\wedge \text{d}\tilde{\text{P}}_{\phi} \wedge \frac{\sqrt{\phi} \ \text{d}\text{C}}{2\big(\text{F}+(2\Lambda^{-1}\partial_{\text{r}}\phi)^2\big)^{\frac{1}{2}}}. 
\ee

Now we need to incorporate $\text{P}_{\text{C}}$, in the measure. From the last equation, in \eqref{cantransojo} we can get explicitly $\text{P}_{\text{C}}$, in terms of the other degrees of freedom. 
\be
\text{P}_{\text{C}}=-\int\limits_{0}^{\text{r}_0}\text{dr} \frac{\Big(\Lambda^2 \text{F}+4(\partial_{\text{r}}\phi)^2\Big)^{\frac{1}{2}}}{2(\text{C}-4\sqrt{\phi})}.
\ee
Its differential is given by 
\be \label{dp}
\text{dP}_{\text{C}}=-\int\limits_{0}^{\text{r}_0}\text{dr} \frac{ \sqrt{\phi} \ \text{d}\Lambda }{2\big(\text{F}+(2\Lambda^{-1}\partial_{\text{r}}\phi)^2\big)^{\frac{1}{2}}}+\ldots \ ,
\ee
where the ellipsis represent the term containing $\text{d}\phi(t)$. 

Now we can form the infinite wedge product $\prod_{t\in[t_i,t_f]}\text{dP}_{\text{C}}(t)$. Note that this product will contain a sum of infinite terms \footnote{For handling the right-hand side of  \eqref{dp} we need to discretize the integral.} but there is one term containing the product over all the points in the bulk of $\text{d}\Lambda(t,r)$. The other terms contain at least one $\text{d}\phi(t,\text{r})$. That is \footnote{Strictly speaking, every term also contains an infinite product of the discretized differentials $\Delta \text{r}$, coming from the discretization of \eqref{dp}. This infinite product can be omitted without loss of generality.} 
\be\label{sum-prod}
\prod_{t\in[t_i,t_f]}\text{dP}_{\text{C}}=-\text{i}\prod_{t\in[t_i,t_f]}\prod_{\text{r}\in[0,\text{r}_0)}\frac{\sqrt{\phi} \ \text{d}\Lambda}{2\big(\text{F}+(2\Lambda^{-1}\partial_{\text{r}}\phi)^2\big)^{\frac{1}{2}}}+\ldots \ ,
\ee
where the $-\text{i}$ factor comes from $(-1)^{\sum 1}=(-1)^{-\frac{1}{2}}$.

The observation here is that upon wedge-multiplicating \eqref{sum-prod} with the infinite wedge product of the differentials  $\text{d}\phi$, the ellipsis contribution cancels out and we are left with 
\be\label{sum-prod1}
\prod_{t\in[t_i,t_f]}\prod_{\text{r}\in[0,\text{r}_0)}\text{dP}_{\text{C}}\wedge \text{d}\phi=-\text{i}\prod_{t\in[t_i,t_f]}\prod_{\text{r}\in[0,\text{r}_0)}\frac{\sqrt{\phi} \ \text{d}\Lambda \wedge \text{d}\phi}{2\big(\text{F}+(2\Lambda^{-1}\partial_{\text{r}}\phi)^2\big)^{\frac{1}{2}}}+\ldots \ ,
\ee

From here it is clear that the measure \eqref{mm2} can be written in terms of $\text{d}\text{P}_{\text{C}}$, as
\be\label{mm3}
\prod_{t\in[t_i,t_f]}\prod_{\text{r}\in[0,\text{r}_0)}\text{H}[\text{h}] \ \text{d}\text{N}\wedge \text{d}\text{N}^{\text{r}} \wedge \text{d}\phi \wedge \text{d}\tilde{\text{P}}_{\phi} \wedge \text{d}\text{C}\wedge \text{d}\text{P}_{\text{C}}. 
\ee 
Note that \eqref{mm3} does not depend on $\text{P}_{\text{C}}$. This fact is what justifies the manipulations to go from equation \eqref{complexaction} to  \eqref{complexaction1}.

\section{Unitarity}{\label{appC}}

One of the  axioms  of quantum mechanics (QM) or quantum field theory (QFT) is: 
The evolution of a closed system is unitary. The state vector$|\psi(t)\rangle$,  at time $t$, is derived from the state vector $|\psi(t_0)\rangle$, at time $t_0$,  by applying a unitary operator $ U (t, t_0)$, called the evolution operator $|\psi(t)\rangle=U (t, t_0)\ |\psi(t_0)\rangle$.

A direct consequence of unitary evolution is the probability $\text{P}(t)$, conservation 
\be 
\text{P}(t)=\langle\psi(t)|\psi(t)\rangle=\langle\psi(t_0)|\psi(t_0)\rangle=\text{P}(t_0).
\ee
In other words, unitary evolution is equivalent to say the probability does not depend on time.

For defining probability in a quantum mechanic's settings there should exist a notion of internal product between states $\langle \cdot | \cdot\rangle$. In the coordinate representation of the wave function, it is defined as 
\be 
\langle\psi(t)|\psi(t)\rangle=\int \text{d}x^\text{d} \text{M(x)} \psi(t,x)\psi^{*}(t,x),
\ee
where $\text{M(x)}$, is a measure.

Over this paper we have been working with the concept of (non) unitary evolution, however, we do not have a clear notion of the product between states.  In what follows we will establish the conditions under which the wave function \eqref{finalamplitudeojo1} evolves unitary even without explicitly knowing the internal product between states.

As seen, there is a possibility of achieving unitary evolution when $\phi_{|_{\text{B}}}\rightarrow \infty$. Let us pose the calculation of the probability for a generic state of the form \eqref{finalamplitudeojo1}
\begin{align}\label{uni}
\text{P}(\tau)=\int \text{D}[\phi_f,\Lambda_f]\Psi\Big[(\phi,\Lambda)_{f},(\phi,\Lambda)_{i}\ ;(\infty,\tau)_{\text{B}}\Big] \Psi^{*}\Big[(\phi,\Lambda)_{f},(\phi,\Lambda)_{i}\ ;(\infty,\tau)_{\text{B}}\Big]= \nonumber \\
\int \text{D}[\phi_f,\Lambda_f]
\int\limits_{0}^{\infty} \text{dC}\  \int\limits_{0}^{\infty} \text{dC}^{\prime}\ \chi(\text{C})  \chi^{*}(\text{C}^{\prime}) 
\text{exp}\Big[\text{i}\Big(\Omega^{\prime}[\Lambda,\phi;\text{C}]-\Omega^{\prime}[\Lambda,\phi;\text{C}^{\prime}]\Big)\Big{|}_{t_i}^{t_f}\Big]\nonumber \\
\text{e}^{-\text{i}\frac{1}{4}\big(\text{C}-\text{C}^{\prime}\big)\tau}. 
\end{align}
Although we do not know the measure $\text{D}[\phi_f,\Lambda_f]$, we can make some conclusions about the time evolution of the probability.

From \eqref{uni} we can see that the condition for the probability does not depend on $\tau$, is
\be\label{uni-ojo-ojo}
\int \text{D}\phi_f \text{D}\Lambda_f \  \text{M}[\phi_f ,\Lambda_f] \
\text{exp}\Big[\text{i}\Big(\Omega^{\prime}[\Lambda_{t_f},\phi_{t_f};\text{C}]-\Omega^{\prime}[\Lambda_{t_f},\phi_{t_f};\text{C}^{\prime}]\Big)\Big]=\text{f}(\text{C})\delta\big(\text{C}-\text{C}^{\prime}\big),
\ee
where $\text{f}(\text{C})$, could be any function. In this case, the probability is 
\be
\text{P}(\tau)=\int\limits_{0}^{\infty} \text{dC}\ \chi(\text{C})  \chi^{*}(\text{C}) \ \text{f}(\text{C})=\text{P}(0). 
\ee

Equation \eqref{uni-ojo-ojo} is extremely difficult to satisfy for some $\text{M}[\phi_f ,\Lambda_f]$. Nonetheless, we do not discard the possibility that there exists a measure such that \eqref{uni-ojo-ojo} holds. We leave this window open and the search for this measure for future works.

 {\it On the one hand, if this measure exists there is only one case where the wave function evolves unitary. It is when $\phi_{|_{\text{B}}}\rightarrow \infty$. On the other hand, if this measure does not exist we inevitably must conclude that time evolution in QG is non-unitary}.

\end{document}